\documentclass[12pt,a4paper,final]{iopart}

\usepackage{iopams}  
\usepackage{graphicx}
\usepackage{float}
\usepackage{caption}
\usepackage{subcaption}
\usepackage[breaklinks=true,colorlinks=true,linkcolor=blue,urlcolor=blue,citecolor=blue]{hyperref}\usepackage{xr}
\expandafter\let\csname equation*\endcsname\relax
\expandafter\let\csname endequation*\endcsname\relax
\usepackage{amsmath}
\usepackage{mathtools}
\usepackage{doi}
\usepackage{xcolor}
\usepackage[
  height=9in,      
  width=7in,       
  top=78pt,        
  headheight=48pt, 
  headsep=12pt,    
  heightrounded,   
  verbose,         
]{geometry}
\usepackage{fancyhdr}

\begin{document}
	\title[ ]{Improving the Segmentation of Scanning Probe Microscope Images using Convolutional Neural Networks}
	
	\author[cor1]{Steff Farley}
	\address{School of Science, Loughborough University, Epinal Way, Loughborough, LE11 3TU, United Kingdom}
	\ead{r.m.farley@lboro.ac.uk}
	
	\author{Jo E.A. Hodgkinson}
	\address{School of Physics \& Astronomy, The University of Nottingham, University Park, Nottingham, NG7 2RD, United Kingdom}
	\ead{Joe.Hodgkinson@nottingham.ac.uk}
	
	\author{Oliver M. Gordon}
	\address{School of Physics \& Astronomy, The University of Nottingham, University Park, Nottingham, NG7 2RD, United Kingdom}
	\ead{Oliver.Gordon@nottingham.ac.uk}
	
	\author{Joanna Turner}
	\address{School of Mechanical, Electrical and Manufacturing Engineering, Loughborough University, Epinal Way, Loughborough, LE11 3TU, United Kingdom}
	\ead{j.turner@lboro.ac.uk}
	
	\author{Andrea Soltoggio}
	\address{School of Science, Loughborough University, Epinal Way, Loughborough, LE11 3TU, United Kingdom}
	\ead{a.soltoggio@lboro.ac.uk}
	
	\author{Philip J. Moriarty}
	\address{School of Physics \& Astronomy, The University of Nottingham, University Park, Nottingham, NG7 2RD, United Kingdom}
	\ead{philip.moriarty@nottingham.ac.uk}
	
	\author{Eugenie Hunsicker}
	\address{School of Science, Loughborough University, Epinal Way, Loughborough, LE11 3TU, United Kingdom}
	\ead{e.hunsicker@lboro.ac.uk}

\begin{abstract}
	A wide range of techniques can be considered for segmentation of images of nanostructured surfaces. Manually segmenting these images is time-consuming and results in a user-dependent segmentation bias, while there is currently no consensus on the best automated segmentation methods for particular techniques, image classes, and samples. Any image segmentation approach must minimise the noise in the images to ensure accurate and meaningful statistical analysis can be carried out. Here we develop protocols for the segmentation of images of 2D assemblies of gold nanoparticles formed on silicon surfaces via deposition from an organic solvent. The evaporation of the solvent drives far-from-equilibrium self-organisation of the particles, producing a wide variety of nano- and micro-structured patterns. We show that a segmentation strategy using the U-Net convolutional neural network outperforms traditional automated approaches and has particular potential in the processing of images of nanostructured systems.
\end{abstract}

\submitto{\MLST}
\vspace{2pc}
\newpage
\section{Introduction}

	Image segmentation is a common step in a workflow to carry out statistical image analysis. There are several techniques that can be employed for image segmentation and deciding upon one can be an important research question that can result in different analysis and conclusions. There is some existing knowledge on the strengths and weakness of various approaches but it is not always theoretically clear which techniques are optimum for particular patterns or noise\cite{Davies2012}. For large datasets, it is also time consuming to choose (and optimise) a method for each image, and so a generalised automated approach is desirable.
	
	Images of nanostructures present a further challenge for image segmentation. The scale of these structures require imaging techniques that are particularly susceptible to surface defects, contamination, and artefacts. When surfaces are imaged using a scanning probe, tip changes can result in loss of resolution, multiple-tip artefacts, discontinuities, and `smearing'. In addition, the properties of the piezoelectric actuators, and the design of the microscope itself, often lead to additional image distortions such as non-linear creep and thermal drift\cite{Straton2014}. This is particularly problematic when comparing images from experimental data to that of simulations produced by computational models which are free of the types of noise and distortion that plague experimental imaging.
	
	In this paper we focus on a broad set of atomic force microscope (AFM) images of 2D nanoparticle assemblies formed on silicon substrates via deposition from a volatile organic solvent. The solvent dewetting process produces a wide variety of nanostrutured patterns\cite{Stannard2011} spanning cellular networks, labyrinthine networks (bearing a striking similarity to those formed in polymer thin films), ensembles of islands, rings, and branched `tree-like' structures, although this is not an exhaustive list. Analysis of the various patterns formed via evaporation of the solvent provides key insights into the dewetting process and enables the development of new protocols for the control of matter at the nanometre scale level\cite{Vlasov2001}\cite{Thiele1998}\cite{Maillard2000}.

	A key issue, however, is that the majority of analysis/classification for nanostructured systems is currently carried out `by hand' due to deficiencies in automated image processing when it comes to handling precisely the type of artefacts that are so prevalent in scanning probe data. Very often, the first step in the analysis of an AFM image of a nanostructured (or microstructured) sample is segmentation in order to locate features. Here we consider traditional automated segmentation approaches and a neural network approach to segmentation and evaluate their ability to handle the types of artefacts found in this data. We aim to provide recommendations for the segmentation process from this evaluation.
	
\section{Dataset and Dataset Curation}

\subsection{Experiment}

	The dataset used contains atomic force microscope (AFM) images of the depositions of gold nanoparticles that have been placed in a solvent and dried on a substrate. These tiff file images are 512x512 pixels in gray-scale format. The parameter space associated with the pattern generation process is very wide and encompasses variables such as the type of solvent, the concentration of nanoparticles, the length of the thiol chain surrounding the Au particle core, the relative humidity, the drying method (e.g. drop or spin cast), and the amount of excess thiol in the nanoparticle solution. See Refs. \cite{PauliacVaujour2007} - \cite{Blunt2007}, for a description of the experimental protocols.

\subsection{Nanostructured Patterns}

\begin{figure}[t]
		\centering
		\includegraphics[width=\textwidth]{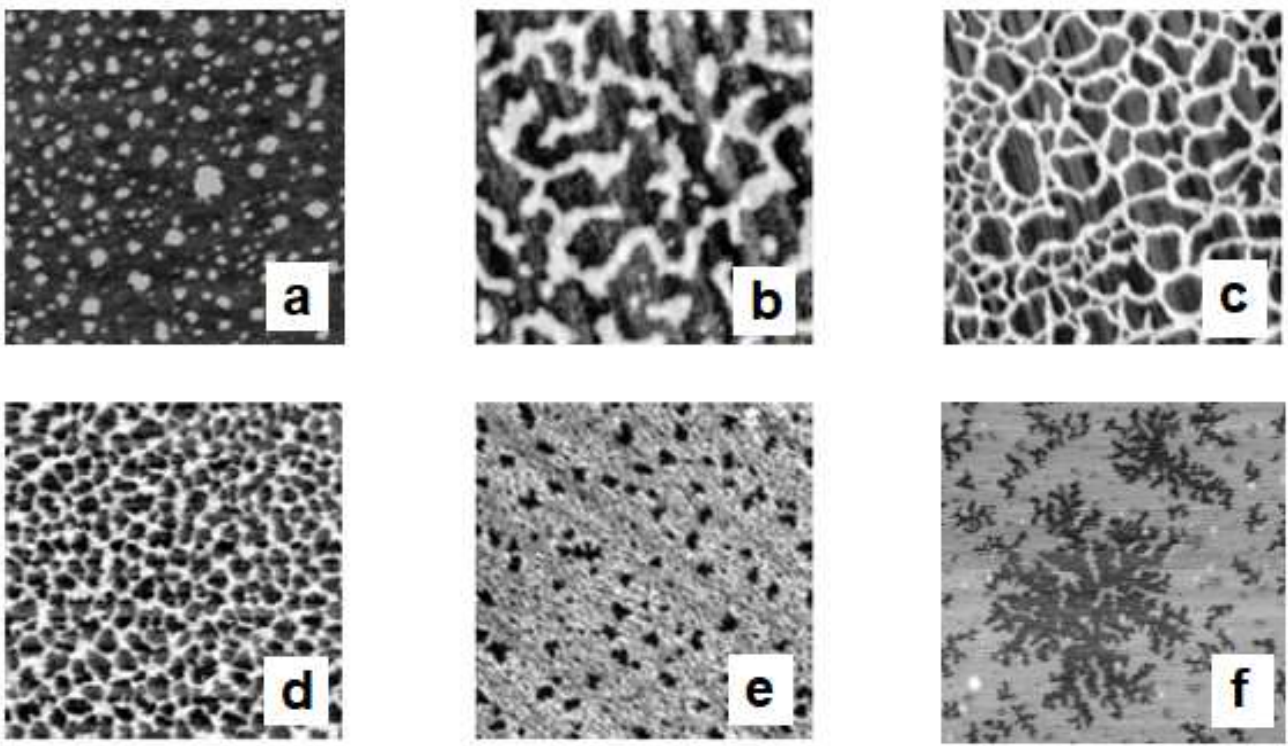}
		\caption{Characteristic examples of a) islands, b) worm-like domains, c) cellular networks, d) labyrinthine, e) porous, and f) fingering regimes identified in literature for AFM nanoparticle experiments.}
		\label{fig:Regimes}
	\end{figure}

	The solvent dewetting process leads to a number of distinct pattern types that have been described in the literature \cite{Stannard2011} as islands, worm-like, cellular, labyrinthine, pores, and fingering patterns. Examples of these regimes are shown in figure \ref{fig:Regimes}. 
	
	Some regimes are qualitatively similar to each other in terms of visual interpretation. Worm-like features could be considered as stretched out islands, and cellular images are similar to connected worm-like features. Labyrinthine structures are similar to cellular image but have less of a tiling pattern. Pores may be seen as more sparsely populated and less ramified labyrinthine structures.  Finally, fingering features could be seen as a variation of pores. There are some images that lie on the border between two of these regimes in this phase diagram. We also see images with two or more regimes represented, as well as multi-scale examples of regimes.
	
\begin{figure}[t]
		\centering
		\includegraphics[width=.8\linewidth]{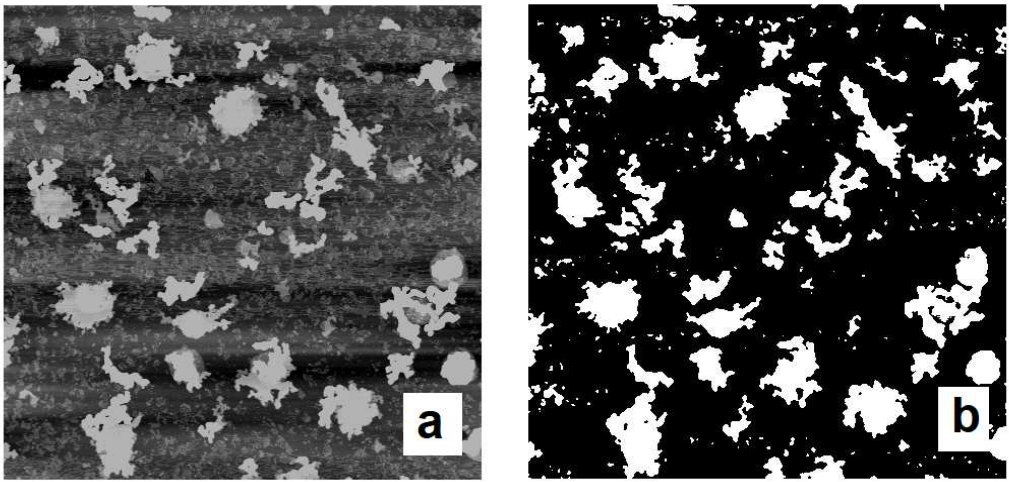}
		\caption{Example of a multilayer image and an attempted segmentation of this image using Otsu's threshold.}
		\label{fig:Multilayer}
	\end{figure}

	We sometimes also observe two or more layers in these images, where an additional layer of nanoparticles has formed, and the pixel intensity of nanoparticles on different layers can vary. An example is shown in figure \ref{fig:Multilayer}. This occurs when the nanoparticle solution has a relatively high concentration\cite{Martin2008}. In this paper we focus on binarisation approaches for single-layer images as multi-layer images require multi-level segmentation approaches which are not evaluated in this paper.
	
\subsection{Pre-processing}

	Scan data returned by scanning probe microscopy can be read into software as gray-scale 2D images. Pixel coordinates represent physical tip position on the scan surface, while pixel intensity values within a single value channel provide representative heights of surface features, providing both effective visual and topographical information. Popular image processing software, such as ImageJ\cite{Schneider2012} and Gwyddion\cite{Necas2012}, focuses on loading individual images, where processing methods and their parameters are controlled by the user based on the appearance of the resulting images. This can lead to biases towards visual information as opposed to complete topographical accuracy. A processed image tolerable by eye for the purpose of presentation often contains bias towards removal of local defects and enhancement of desired surface features, resulting in an image that may still have large quantities of noise that will reduce the quality of its future segmentation. Alternatively, an automated pre-processing regime is less time-consuming than manual processing within a popular image processing software, and promotes consistency in image statistics. Pre-processing software using a combination of python and R language code was adopted.
 
	Raw data obtained directly during routine SPM experimentation will contain artifacts produced by the scanning probe. Line-by-line acquisition results in ubiquitous scan line artifacts, leading to incongruity particularly visible in multi-layer images, while raw output data is not systematically leveled which significantly affects the quality of binarisation. The proposed pre-processing software attempts to remove or reduce scanning artefacts using two algorithms to return a data-leveled image with spatially-continuous layers. A corrective line-by-line alignment algorithm is deployed along the same axis as line-by-line scans. Pixel values in each row are shifted to have the median difference between pixels in neighbouring rows in the nearest row equal zero. Median of differences row alignment was found to be the most effective technique for creating continuous features but does not sufficiently level the image. An orthogonal polynomial detrending algorithm with three degrees of freedom returned sufficiently leveled images; the low degree of freedom results in a general image background subtraction without data loss due to equations fitting to data instead of noise. Subtle or local fluctuations in images identified by higher degrees of freedom fits heavily diminished during binarisation.
	
\subsubsection{Contrast Normalisation}

	The contrast of these images is then normalised such that the minimum pixel value becomes 0 and the maximum value becomes 255, using a combination of shifts and scaling operations\cite{Bradski2000}.

	In some images, the minimum and maximum pixel intensities are already at these values. Therefore pixels outside a certain distance from the mean are identified and we then apply contrast normalisation excluding these pixels, which are then set to 0 or 255 by default depending on whether they were smaller or larger than the mean. We consider using 3, 2, or 1 standard deviations from the mean and evaluate which option provides the best segmentation results.
	
\subsection{Noise}

\begin{figure}[t]
		\centering
		\includegraphics[width=\textwidth]{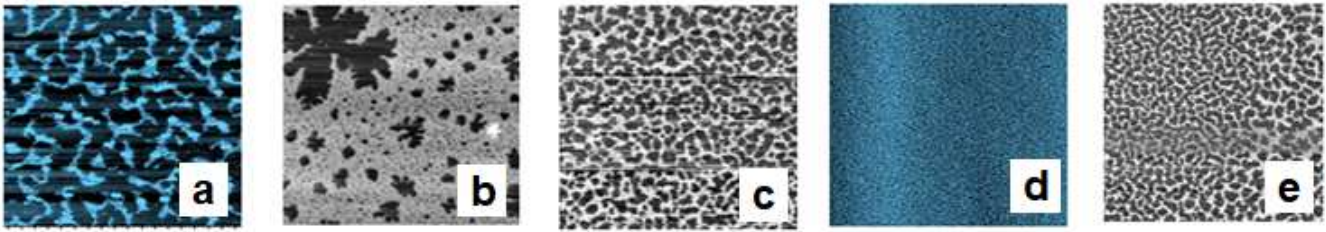}
		\caption{Examples of a) streaking, b) background contrast, c) striping, d) banding, and e) blurring noise found in this dataset}
		\label{fig:Noises}
	\end{figure}

	The common types of noise we encountered can be described as banding, streaking, background contrast, stripes, and blurring. Examples of each of these are given in figure \ref{fig:Noises}. These types of noise have been observed and discussed in previous literature on SPM images\cite{Stirling2013}. We can also see the relevance of such noise by the fact that packages with image processing functions have been developed, such as Gwyddion\cite{Necas2012}, to deal with these types of noise, albeit not in an automated manner.
	
	The most problematic noise types were banding, background contrast and streaking, as we observed that these had the greatest impact on statistics extracted from the images when they were artificially added to relatively clean examples. We demonstrated this by selecting 28 images that were accurately segmented using two common methods: Otsu's threshold\cite{Otsu1979} and a local mean threshold. We then artificially added the five most common noise types to these images. Using the same segmentation methods, we binarised these new noise augmented images and extracted common statistics, observing how these these statistics changed in noise-augmented images from the originals. We used the 2D Minkowski numbers - perimeter, area, and Euler characteristic - as statistical measures. These have been used previously to analyse the distribution of features in these images\cite{Martin2005}\cite{Stannard2008}. For the Euler characteristic, we measured the average difference between noise augmented images and the originals. For area and perimeter we measured the average difference relative to the mean. The largest average differences were seen for background contrast, banding, and streaking, suggesting that these noise types have the greatest impact on accurate segmentation. See table \ref{tab:noiseMinks} in the Supplementary Information for the full results of this experimentation.
	
\subsection{Manual Image Segmentation of Data}

	Developing a generalised, accurate and repeatable segmentation method is an important task as small changes in threshold can lead to different segmented images and subsequently substantially different extracted statistics.
	
\begin{figure}[t]
		\centering
		\includegraphics[width=\textwidth]{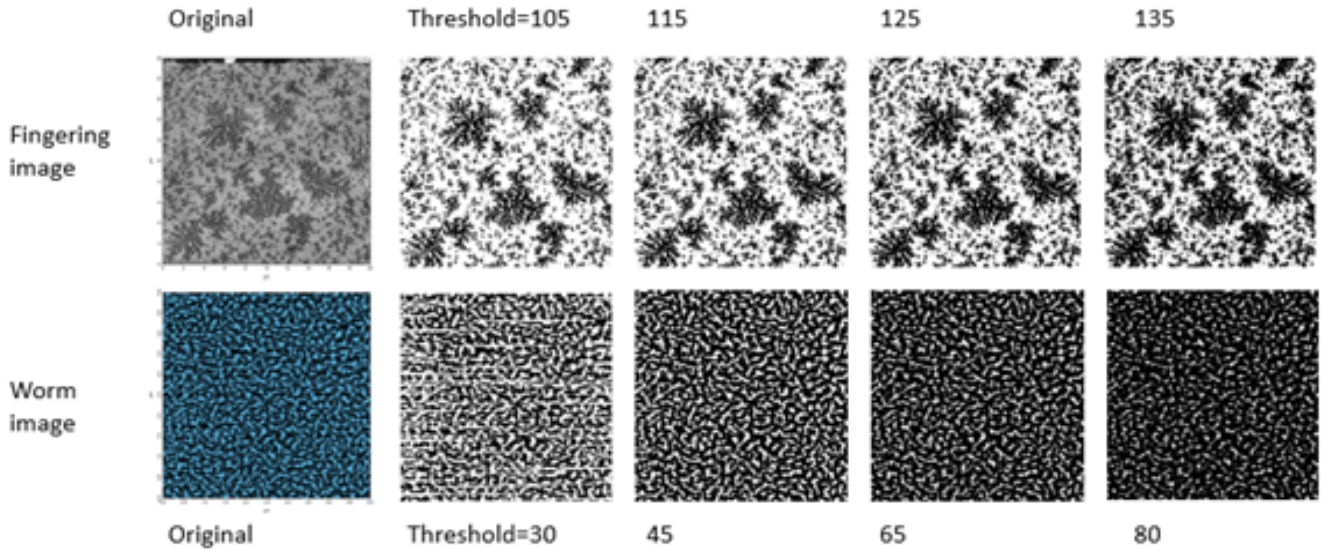}
		\caption{Images taken at a series of thresholds that cover the range of threshold values subjectively chosen by seven individuals: 105, 115, 125, and 135 for a fingering image and 30, 45, 65, and 80 for a worm-like image.}
		\label{fig:ManualThresh}
	\end{figure}

	To demonstrate this, we took one characteristic example of a fingering regime and one characteristic example of a worm-like regime and segmented at four different thresholds for each image. By taking four different thresholds we can observe the variability of common statistics that can be extracted from these images and used in metrics. These images were chosen because fingering and worm-like images are qualitatively different. The original images are shown in figure \ref{fig:ManualThresh}.
	
	The four different thresholds were chosen to cover a range of threshold values such that they spanned the range of thresholds manually chosen by seven individuals, provided in table \ref{tab:ManThreshes} the Supplementary Information. These individuals were asked to set a threshold that they believed gave the best binarisation for these images. From the images in figure \ref{fig:ManualThresh} we can see that the visual difference between these images is not substantial but normalised Minkowski numbers extracted from each of these images showed a substantial variation across the thresholds.  See table \ref{tab:threshMinks} in the Supplementary Information for the full results of this experimentation. The worm-like image was particularly sensitive to the choice of threshold. This shows us that repeatability, along with accuracy, is an important quality for any segmentation approach.
	
\subsection{Dataset Curation}

	To evaluate segmentation approaches, we need a dataset that is representative of the structures and noise that are observed in these images. For training a network for semantic segmentation, we need several examples of each structure and noise type to construct a training dataset that produces a network with sufficient accuracy. For each training image, we produce a corresponding segmented image from either manual segmentation or from using an automated segmentation method with sufficiently accurate results by visual inspection. We also need a representative testing dataset to evaluate the segmentation approaches.
	
\subsubsection{Inclusion Criteria}

	The dataset consisted originally of 2,625 images, taken at various scales.  Images from the six regimes identified above from the literature were first identified and included in the study set. From the remaining images, other regimes that appeared less frequently in the dataset were identified, including rings, trees, streaks, and cracks. All images from those regimes were added to the new dataset. Images with two or more regimes present were also included, as well as images with multiple layers. For those with multiple layers, we segmented the most prominent layer. For each image, the regime and noise were manually labelled by visual observations.
	
\subsubsection{Exclusion Criteria}

	Those with excessive noise or no clear regime were not included in either the training or testing set. An example of an image with excessive noise is given in figure \ref{fig:wownoisy} in the supplementary information. This left 728 remaining images in the dataset.
	
\subsubsection{Training and Testing Set Separation}

	This set was separated into training and testing sets by choosing approximately 75\% of images from each domain to be training data, stratifying by noise type. This was achieved by assigning a random number between 0 and 1 to each image and selecting approximately the first 25\% of images with the greatest number from each domain and noise type for the testing set. When there was only a single case of a particular noise type for a particular domain, that image was included in the testing set if the random number associated with it was greater than 0.5.
	
	We therefore produced an original training dataset of size 428 and a testing dataset of size 300. Each image has the following fields associated: regime, noise, scale and multi-layer indicator flag. A breakdown of these datasets by regime and noise type can be found in table \ref{tab:breakdown} in the Supplementary Information.
	
\subsubsection{Augmentation}

\begin{figure}[t]
		\centering
		\includegraphics[width=\textwidth]{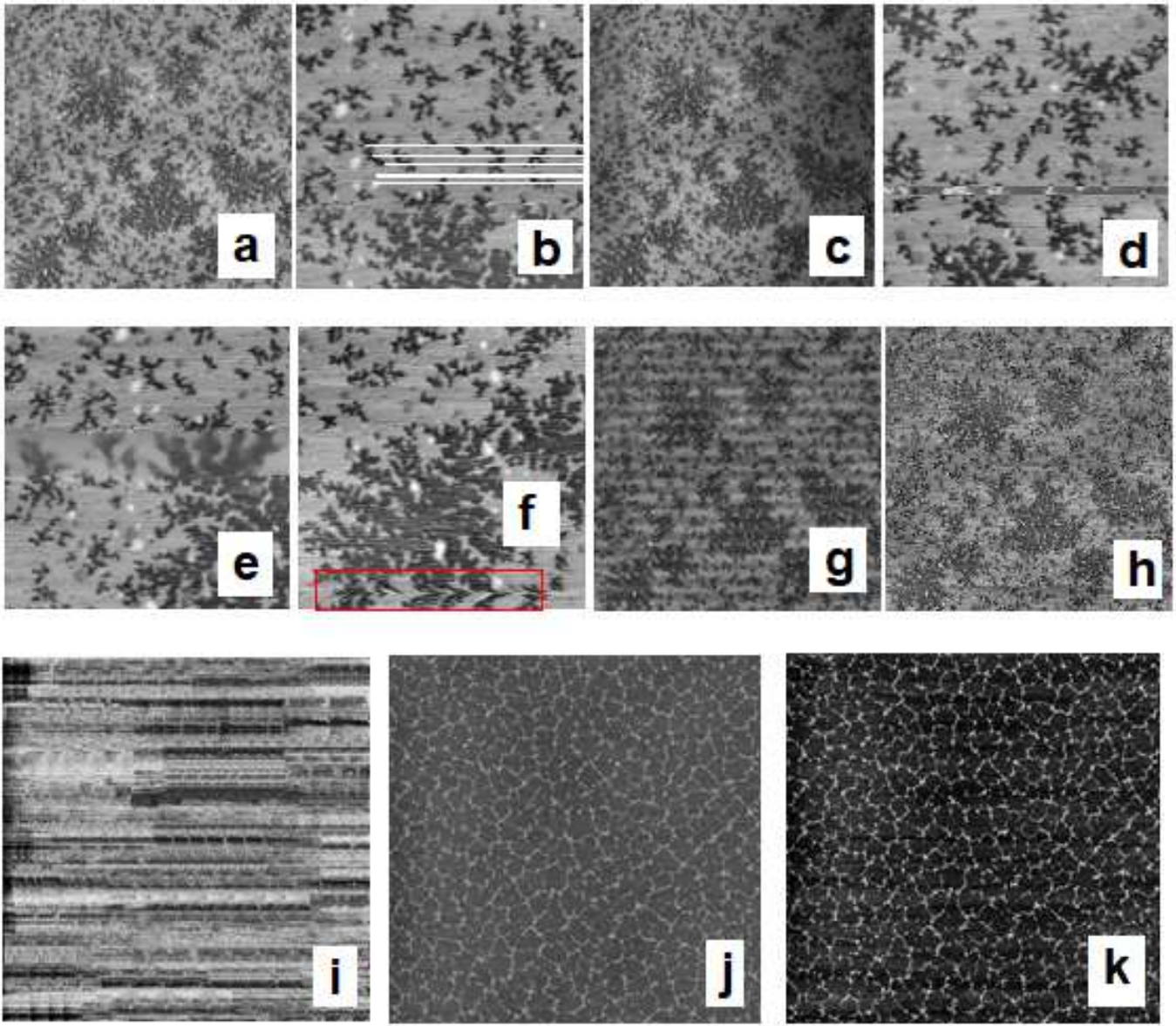}
		\caption{Examples from artificial noise augmented process 2 to clean images. A) Original image, b) stripes, c) vertical banding, d) inversion, e) blurring, f) drift (in red box), g) streaking with high frequency horizontal banding, h) varying background contrast. Image i) shows a mask and k) demonstrates this mask applied to image j) to imitate streaking.}
		\label{fig:Augmentation}
	\end{figure}

	To ensure that the dataset was representative of all noise types for all regimes, we augmented a subset of 80 originally clean images with artificial noise. We consider three noise augmentation processes. For Noise Augmentation Process 1, the types of noise that we introduced were multiple stripes, banding, streaking, background contrast, inversion, blurring, drift, and a combination of banding and stripes. This produced 560 additional images. Examples of an original image and added noise are shown in figure \ref{fig:Augmentation}. Streaking was created by applying a mask to black pixels the image. The inverse of the mask was also applied to white pixels in images to create an augmented image with varying background contrast. This mask could be augmented with flips and scales to vary the streaking between images. The mask, shown in figure \ref{fig:Augmentation}, was taken from a streaking image from the training dataset.

	As we already have the manual segmentations of the clean images that we were artificially adding noise to, we were able to use the same segmented images for training networks for both the clean and noisy examples.
	
	For Noise Augmentation Process 2, we introduced the same noise as in Process 1 but replaced the streaking mask with a high frequency horizontal banding, which also resulted in the same number of additional images for training (560). For Noise Augmentation Process 3, we removed uncommon noise types and only augmented with stripes, banding, streaking with mask, background contrast, and a combination of banding and stripes. Process 3 provides us with 400 augmented images.
	
\section{Methodology}

	In this work we have reviewed several binarisation approaches in combination with pre-processing operations. A complete list of pre-processing, binarisation and post-processing approaches considered is given in table \ref{tab:bin_approaches}.
	
	\begin{table}[H]
		\begin{center}
		\caption{The pre-processing, binarisation and post-processing approaches that were reviewed.}
		\label{tab:bin_approaches} 
		\begin{tabular}{|c|c|c|}
		\hline
		{\bf Pre-processing} & {\bf Binarisation} & {\bf Post-processing} \\
		\hline
		K-means, $K>2$ & Global mean threshold & Despeckling \\
		\hline
		Mean shift & Local mean threshold & Autoencoder \\
		\hline
		Gaussian filter & Global Otsu's threshold & \\
		\hline
		Histogram equalisation & U-Net & \\
		\hline
		\end{tabular}
		\end{center}
	\end{table}
	
\begin{figure}[t]
		\centering
		\includegraphics[width=\textwidth]{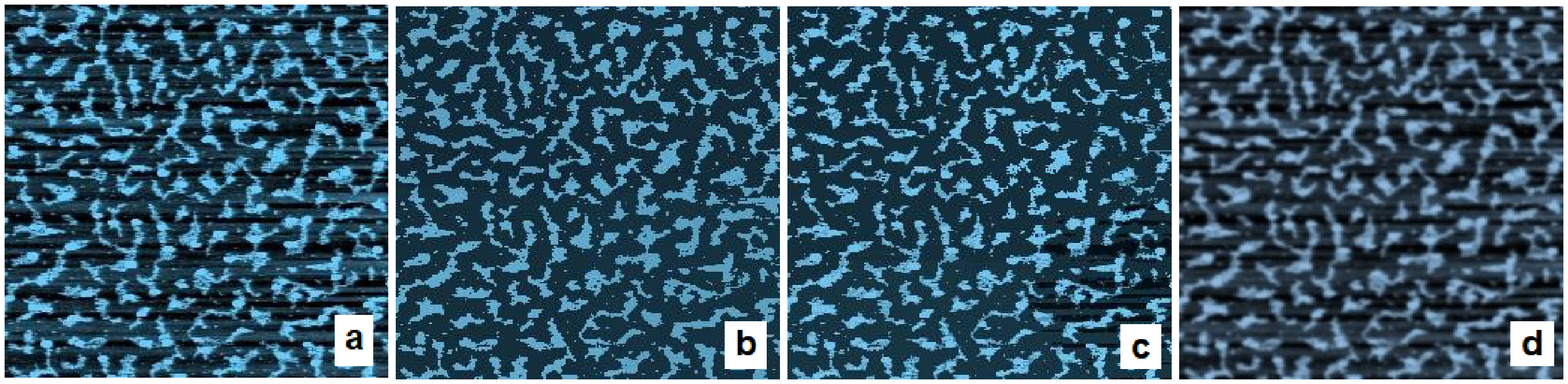}
		\caption{a) Original image, b) image with K-means segmentation, k=8, c) image with mean shift segmentation, d) image with Gaussian filter.}
		\label{fig:Preprocessing}
	\end{figure}

	K-means clustering with $K>2$ carries out multi-class image segmentation but can be used as a pre-processing operation to deal with some noise in images. For each pixel, it uses the three colour channels and the two pixel coordinates as the input data to be assigned to clusters. This clustering approach therefore takes into account both colour and spatial information. We used code written by A. Asvadi\cite{Asvadi2015} to implement this, which uses the Euclidean distance between data points and cluster centroids. An image from the dataset and the same image after k-means clustering with $K=8$ are shown in figure \ref{fig:Preprocessing}.
	
	We also used code by A. Asvadi\cite{Asvadi2015} to implement mean shift clustering, which is another multi-class image segmentation approach that can also deal with noise. This is a non-parametric clustering technique that does not require the number of clusters to be specified. It starts by using a single random point in the data as an initial `mean' point $m_1$ and it then looks at the new mean $m_2$ of all points in a window with a set bandwidth $b$ around the previous mean. If the Euclidean distance between the new and old means $d_{(m_1,m_2)}$ is greater than $\frac{b}{2}$, then a new cluster is created, with the exception being that a new cluster is always created on the first iteration even if this criteria is not met, so that there is always at least two clusters. The process continues iteratively until $d_{(m_i,m_{i+1})}$ is below a set threshold. Data points are assigned to the cluster that they were within the bandwidth of most frequently across all iterations. In our case, the data points were again the three colour channels and two pixel coordinates for each pixel in an image. The original image from figure \ref{fig:Preprocessing} is also shown with mean shift clustering applied to it.
	
	A Gaussian filter can remove Gaussian noise in an image by convoluting the image with a Gaussian kernel\cite{Leach2010}. This is a filter that is commonly applied to noisy images, particularly those with salt and pepper noise. To apply this filter, we specify the size of the kernel and the standard deviation. This results in a slightly blurred image, dealing with noise but also reducing the quality of the image, as seen in figure \ref{fig:Preprocessing}. We aim to evaluate whether this improves the overall quality of the binary image after applying binarisation methods.
	
	The last pre-processing approach we considered was histogram equalisation. This is a common technique to normalise the contrast in the images before binarisation techniques are applied by attempting the create a uniform distribution of pixel intensities in the image\cite{McReynolds2005}. We have evaluated this pre-processing technique in combination with the others.
	
	For binarisation approaches, we can first consider a simple thresholding approach that uses the mean of all pixel intensities in the image as the threshold value. Any pixels above this value are set to white and any below this value are set to black. An adaptive mean thresholding technique uses a local mean in a moving window within the image, such that each pixel has a different threshold value determined by the mean pixel intensity in the window surrounding it\cite{Dey2014}.
	
	A more sophisticated global approach to thresholding is Otsu's threshold\cite{Otsu1979}. This chooses a threshold value such that the variation within the two groups (pixels set to black and pixels set to white) is minimised, or rather, the inter-class variance is maximised. This considers more information about the pixel intensity distribution and as such is expected to perform better than the global mean threshold. However, it assumes that the distribution of recorded scan heights is bi-modal.
	
	U-Net is a convolutional neural network architecture designed for segmenting images, where the task is to predict a classification for each pixel\cite{Zhixuhao2017}. This architecture has been shown to achieve state-of-the-art performance for biomedical image segmentation\cite{Ronneberger2015}. Its unique U-shaped architecture, made up of a contracting path and an expanding path, facilitates the use of contextual information to inform pixel predictions, and precise localisation.

	The task of nanoparticle deposition segmentation has similar requirements to biomedical segmentation, it was therefore hypothesised that the U-Net architecture could be well suited for this problem. Precise localisation is required to infer accurate and meaningful statistics. Contextual information in the image can provide clues about the type of regime and therefore may assist with the predictions of class for each pixel. The dataset we curated was used to train this network. The U-Net training strategy can address the need to learn invariance to challenging noise types, such as banding, contrast and streaking, with only few labelled images for training (428).

	We consider this network with and without the noise augmented images included in the training dataset. We also considered the different noise augmentation options, different pixel histogram truncation before contrast normalisation, and the order in which we performed noise augmentation and contrast normalisation. We evaluated several models but in this paper we focus on the two best performing configurations, that we named Unet1 and Unet2. See table \ref{tab:unetsettings} in the Supplementary Information for details on the settings used for training these U-Net models.
	
	\begin{table}[H]
		\begin{center}
		\caption{U-Net model trained with different settings.}
		\label{tab:unetmodels} 
		\begin{tabular}{|c|c|c|}
		\hline
		{\bf Name} & {\bf Noise augmentation} & {\bf Contrast normalisation} \\
		\hline
		Unet1 & Noise augmentation type 1 & 2 standard deviations  \\
		\hline
		Unet2 & Noise augmentation type 3 & 2 standard deviations  \\
		\hline
		\end{tabular}
		\end{center}
	\end{table}

	Despeckling is a post-processing technique replaces each pixel with the median value within its 3x3 pixel neighbourhood\cite{Schneider2012}. This is effective at removing `salt and pepper' noise, that is, small enclosed regions of black or white pixels.
	
	The second post-processing method we consider is a denoising autoencoder. This has been successfully employed to improve classification of AFM images of nanoparticle depositions\cite{Gordon2020}. Input images are encoded and then decoded by an autoencoder, before the reproduced, original image is outputted. The autoencoder developed and described in Gordon et al.\cite{Gordon2020} was trained on simulated images with artificial speckle noise and taught to output the original image without this noise.
	
\section{Results}

\subsection{Qualitative Analysis}

	We can make some simple initial observations through evaluating the identified segmentation approaches visually on the 300 single-layer testing images that were representative of all regimes and noise types.
	
\subsubsection{Pre-processing}

	The first result we observed was that histogram equalisation in combination with any other pre-processing method and any binarisation approached performed poorly, created more noise than the same approaches without histogram equalisation and often produced completely unusable binarised images.
	
	We also found that using k-means or mean shift resulted in information lost before image segmentation without dealing with noise. Applying a Gaussian filter also rarely improved segmentation and typically lead to coarse features. Using no additional pre-processing to alignment, detrending, and contrast normalisation, typically produced the best results.
	
	\begin{figure}[t]
		\centering
		\includegraphics[width=.8\linewidth]{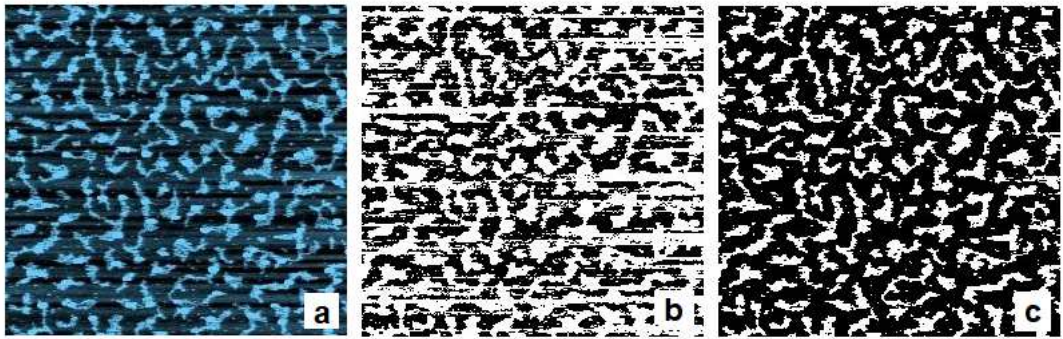}
		\caption{An image segmented by local mean where b) histogram equalisation was applied prior to segmentation and c) where histogram equalisation was not applied.}
		\label{fig:HistEq}
	\end{figure}
	
\subsubsection{Image segmentation}

\begin{figure}[t]
		\centering
		\includegraphics[width=\textwidth]{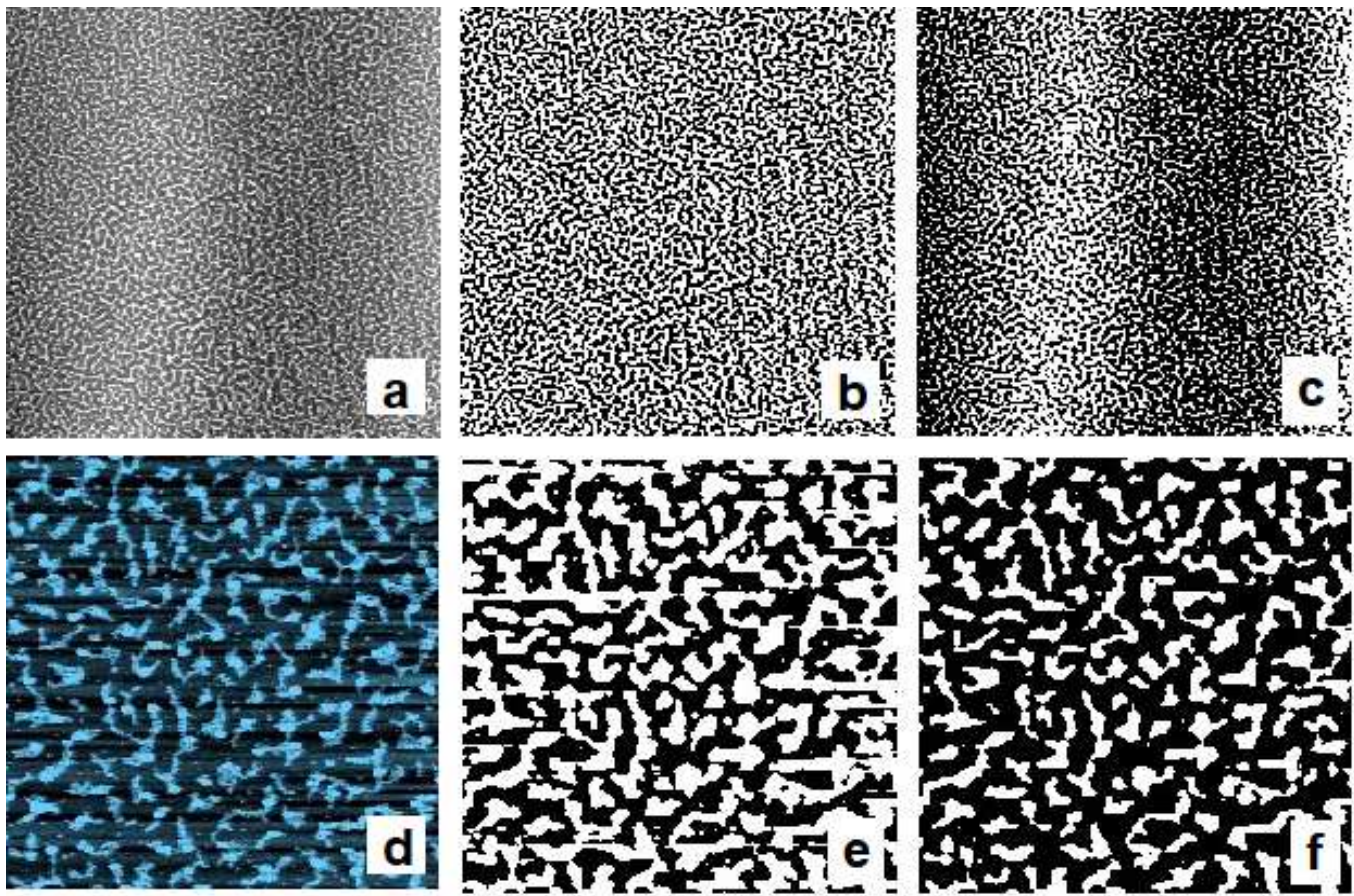}
		\caption{The top row shows a) an image with banding segmented with b) local mean and c) Otsu's threshold. The bottom row shows d) an image with streaking segmented with e) local mean and f) Otsu's threshold.}
		\label{fig:OtsuLM}
	\end{figure}

	As there were different types of noise that appeared in different images, no single option appeared to be able to deal with all of them optimally and some options appeared to perform better on certain types of noise and other options were more suited to different types of noise. For example, figure \ref{fig:OtsuLM} shows how using a local mean threshold deals well with banding in the cellular image where Otsu's threshold still contains substantial banding, however, the local mean approach contains streaks in the worm-like image, which do not appear when using Otsu's threshold. This is a trend we observed throughout the testing dataset. Where Otsu's threshold performed best for streaking, it was also usually improved by applying despeckling too.
	
\begin{figure}[t]
		\centering
		\includegraphics[width=.8\linewidth]{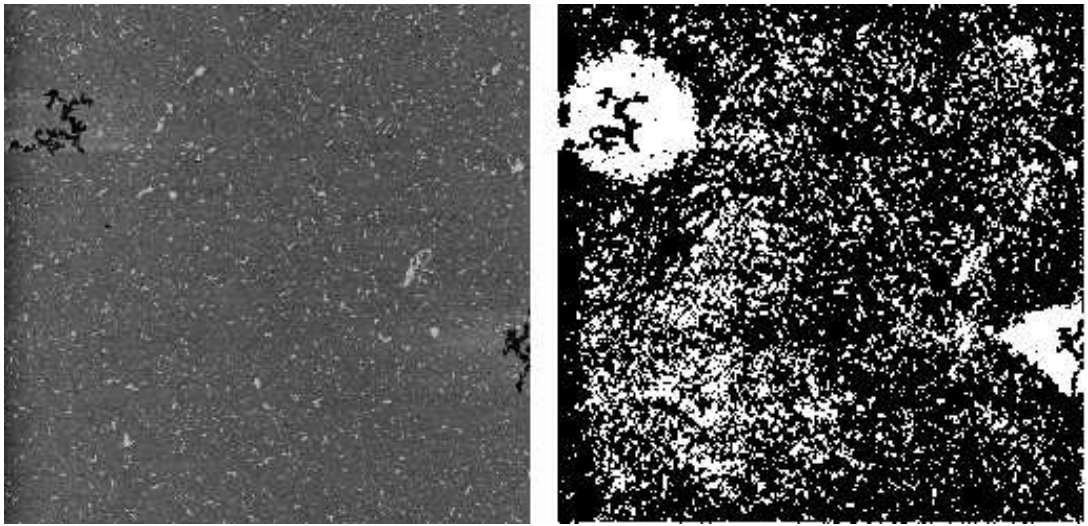}
		\caption{Example of a multilayer image segmented using a U-Net model.}
		\label{fig:UNetMultilayer}
	\end{figure}

	We then evaluated the performance of the U-Net networks for segmentation. We note that the U-Net models trained on two class segmentations fail with multilayer images, as seen in the example in figure \ref{fig:UNetMultilayer}. This is because the model has been optimised for single-layer images, which have been the focus of this paper, and the U-net model can only be evaluated on multilayer images when further work has been done to train the U-net model on these images. The only approach we evaluated that deals with multilayer images well is Otsu's threshold. We therefore only evaluated the binarisation approaches on the remaining 107 single-layer testing images.
	
	We also notice that Unet1 and Unet2, which were trained on more complex noise augmentation and contrast normalisation, were significantly better than a basic U-Net model trained on no noise augmented data.
	
	We compared the segmentations of the 107 single-layer testing images for Unet1 and Unet2 and observed that Unet1 appeared to perform better on streaking and contrast noise, whereas Unet2 appeared to deal with banding noise better, but both U-Net models handled all of these different noise types reasonably well.
	
	For the 107 single-layer testing images, Unet1 performed best according to subjective manual observation in 62 images, Unet2 performed best in 60, Otsu's threshold with despeckling performed best in 65, and local mean with despeckling performed best in 56, where multiple approaches can perform equally well for the same image. The methods performed similarly in the total number of images they were able to give the best segmentation for but this changed for different noise types. Moreover, the U-Net models had a tendency to give reasonably good segmentations for images that they did not perform best for, and there were very few images where they substantially failed to give a good segmentation. This is contrary to the Otsu's threshold and local mean methods, that often failed completely for banding and streaking noise respectively. Table \ref{tab:subjectiveresults} shows how each method performs for different noise types.
	
	\begin{figure}[t]
		\centering
		\includegraphics[width=\textwidth]{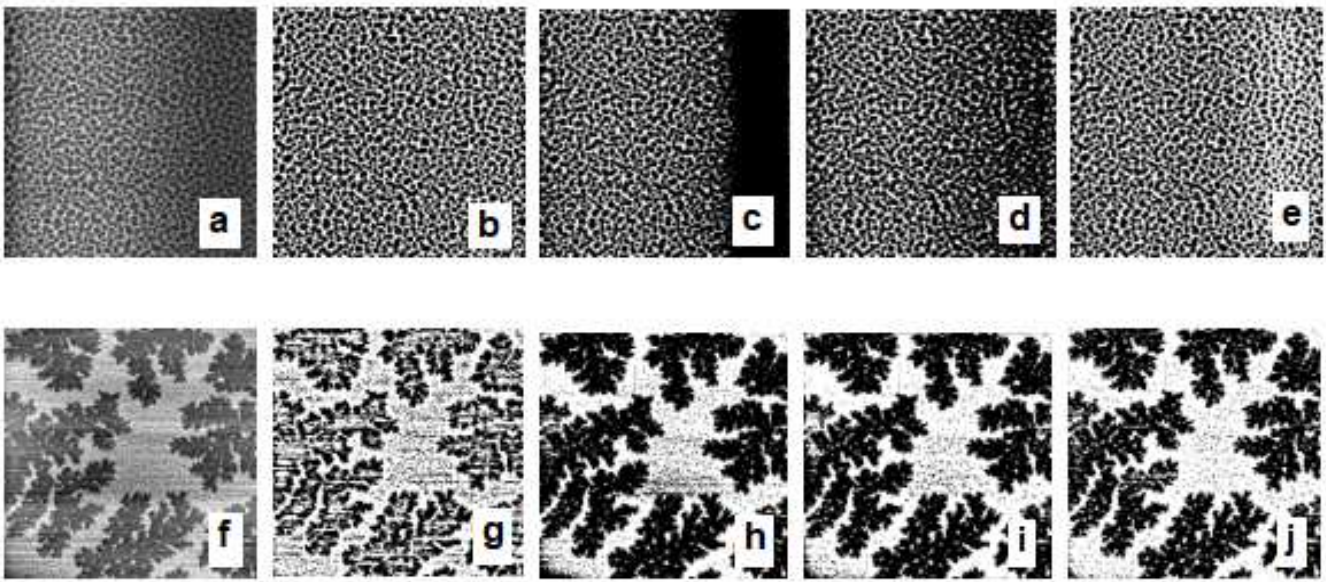}
		\caption{a) Original image with banding, b) image segmented by local mean, c) by Otsu's threshold d) by Unet1, e) by Unet2 f) Original image with streaking and varying background contrast, g) image segmented by local mean, h) by Otsu's threshold, i) by Unet1, j) by Unet2}
		\label{fig:Segmentations}
	\end{figure}
	
	\begin{table}[H]
		\begin{center}
		\caption{Number of testing images with various noise types that was best segmented by each of four methods.}
		\label{tab:subjectiveresults} 
		\begin{tabular}{|c|c|c|c|c|c|}	
		\hline
		& {\bf Local mean} & {\bf Otsu's} & {\bf Unet1} & {\bf Unet2} & {\bf Number of} \\
		& & {\bf threshold} & & & {\bf images} \\
		\hline
		{\bf Background contrast,} & 30 & 43 & 42 & 30 & 65 \\
		{\bf streaking and/or stripes only} & & & & & \\
		\hline
		{\bf Banding only or } & 10 & 8 & 6 & 14 & 21 \\
		{\bf banding with stripes only} & & & & & \\
		\hline
		{\bf Blurring} & 5 & 7 & 7 & 7 & 11 \\
		\hline
		\end{tabular}
		\end{center}
	\end{table}
	
	In testing images that contained only streaking, contrast or stripes and no other noise types, Unet1 and Otsu's threshold appeared to perform best according to subjective visual evaluation of the images. In images that only contained banding or banding and stripes, Unet2 and local mean with despeckling both performed best.

	These results show that Unet1 appears to perform similarly to Otsu's threshold in images that lend themselves to the strengths of Otsu's threshold, while Unet1 also typically outperformed Otsu's threshold for images containing banding. While Unet1 is weaker at banding than local mean, it is able to deal with streaking noise much better. Unet2 performed substantially better than Otsu's threshold at banding and gave better results than local mean for streaking. This provides some evidence that it is possible to train a U-Net model for a more generalised approach to segmentation of these images. Additional experimentation with the training dataset and the settings of the network could improve on these results, finding a balance between the strengths of Unet1 and Unet2, and providing a more robust and generalised segmentation method.
	
	The denoising autoencoder was also applied to segmentations from this dataset in Gordon et al.\cite{Gordon2020} and compared to the same segmentations without the autoencoder to determine whether this resulted in improvements to the segmentations for purpose of image classification. They found that it had some success at denoising striping and streaking artefacts. As the autoencoder was trained to remove speckle noise, it was expected that it would perform well on the thin artefacts associated with striping and streaking noise. These results suggest that applying an autoencoder as part of post-processing improves the results of segmentation for methods that had relatively greater difficulty with streaking, such as the U-Net models.

\subsection{Quantitative Analysis}

	The performance of each segmentation methods were compared quantitatively by observing the proportion of pixels that changed in 16 segmented noise augmented images from their segmented originals for each method. We selected 16 images that each segmentation method was capable of segmenting well and artificially added banding, blurring and striping noise to these examples. Using the same segmentation methods, we binarised these new noise augmented images and determined how many pixels changed (from black to white or from white to black) in the segmented noise augmented images from the segmented original clean images. We take the average proportion of pixels that changed in the 16 images for each artificial noise type and each segmentation method.

	\begin{table}[H]
		\begin{center}
		\caption{Average proportion of pixels changed when banding, blurring, and stripes are artificially added to images and segmented.}
		\label{tab:pixelchanges1} 
		\begin{tabular}{|c|c|c|c|c|}	
		\hline
		& {\bf Local mean} & {\bf Otsu's threshold} & {\bf Unet1} & {\bf Unet2} \\
		\hline
		{\bf Banding} & 1.4\% & 15.9\% & 12.2\% & 6.2\% \\
		\hline
		{\bf Blurring} & 0.2\% & 0.2\% & 0.3\% & 0.4\% \\
		\hline
		{\bf Stripes} & 0.3\% & 0.3\% & 0.5\% & 0.4\% \\
		\hline
		\end{tabular}
		\end{center}
	\end{table}
	
	As expected, the U-Net models both showed improvement on handling banding noise as compared to the values from Otsu's threshold but neither were able to match the performance of local mean for this type of noise.
	
	The features in these images were relatively small as we chose images for which the segmentation methods perform well and local mean struggles to correctly segment larger features even in the absence of noise. Streaking and varying background contrast have less of an impact on segmentation when features are small so to evaluate the methods on these noise types, we split each image into four quarters and rescaled each sub-image to 512x512 pixels to increase the size of the features. This created 4 images from each original image, providing a total of 64 images for this analysis. We artificially added varying background contrast and streaking to these images and segmented them using the four methods we are evaluating. We again determined how many pixels changed from these images to the same sub-images (rescaled) from the original segmentations with no artificial noise.
	
	\begin{table}[H]
		\begin{center}
		\caption{Average proportion of pixels changed when varying background contrast and streaking are artificially added to images and segmented.}
		\label{tab:pixelchanges2} 
		\begin{tabular}{|c|c|c|c|c|}	
		\hline
		& {\bf Local mean} & {\bf Otsu's threshold} & {\bf Unet1} & {\bf Unet2} \\
		\hline
		{\bf Background contrast} & 10.7\% & 4.7\% & 8\% & 10.3\% \\
		\hline
		{\bf Streaking} & 10.6\% & 4.7\% & 7.3\% & 9.6\% \\
		\hline
		\end{tabular}
		\end{center}
	\end{table}
	
	We found that Unet1 handled both varying background contrast and streaking better than local mean but was not able to match the performance of Otsu's threshold for these noise types. This analysis produced results that match the manual observations from the qualitative analysis.
	
	We additionally measured the effect of applying post-processing to the Unet1 segmentation and compared the performance of despeckling and the denoising autoencoder as post-processing methods. We found that these methods performed similarly and, while both proved capable of removing artificial streaking and striping noise, both methods also removed features in the images that were not noise artefacts. These results are shown in table \ref{tab:postproc}. There is a trade-off as both methods remove noise and some features but, depending on the application and the goals of segmentation, this post-processing may be desirable. The autoencoder has shown to be favourable for improving classification rate\cite{Gordon2020}, for example, but using the autoencoder or despeckling would not be desirable if the goal of segmentation is to study the fine structure of the material.
	
	\begin{table}[H]
		\begin{center}
		\caption{Average proportion of pixels changed when noise is artificially added to images and segmented by Unet1 with different post-processing methods.}
		\label{tab:postproc} 
		\begin{tabular}{|c|c|c|}	
		\hline
		& {\bf Autoencoder} & {\bf Despeckle}  \\
		\hline
		{\bf Background contrast} & 7.6\% & 6.6\% \\
		\hline
		{\bf Banding} & 13.7\% & 13.4\% \\
		\hline
		{\bf Blurring} & 3.7\% & 2.5\% \\
		\hline
		{\bf Streaking} & 6.9\% & 6.5\% \\
		\hline
		{\bf Stripes} & 3.9\% & 2.7\% \\
		\hline
		\end{tabular}
		\end{center}
	\end{table}

\section{Conclusion}

	We have investigated a machine learning approach for binary segmentation of noise images and developed an approach that performs as well as traditional methods where they succeed and outperforms on images where each traditional method fails. The evaluation of approaches showed that local mean and Otsu's threshold behave very differently and perform best on different images, whereas the network we have developed incorporated the best of both of these approaches. The U-Net models we developed do not always necessarily provide the best segmentations but typically outperform either Otsu's threshold or local mean, depending on the type of noise present, as shown in figure \ref{fig:Segmentations}.
	
	Through some modification to the size of the training dataset, the addition of augmented noisy images, and changes to the contrast normalisation, we explored ways of improving the network. This showed that some choices in training dataset curation can significantly alter the performance of the network and further attention to this may lead to greater improvements. Further improvements may also be found through altering the settings and structure of the network. The results of this research show that this would be worthwhile endeavour.
	
	While the U-Net network did not perform well for multilayer images, a different training dataset with multiple labels to annotate different layers may be used to similarly develop a network that can handle the multi-classification problem of multiple layers. The results for single-layer images provides optimism for the use of a machine learning approach for multilayer images.
	
	There was no single general method that worked best for all images, so future research may also include a machine learning approach to recognise noise in images and automate the identification of the most suitable segmentation method.

\section*{Acknowledgements}
	
	The authors gratefully acknowledge Emmanuelle Pauliac-Vaujour who also produced the dataset of SPM images used in this work.

\section*{Supporting Information Available}
	
	\begin{itemize}	
	\item A complete dataset of all pre-processed images from the training and testing datasets, the training labels, and the Unet1 model are openly available at \url{https://doi.org10.17639/nott.7072}.
	
	\item Code is available at \url{https://github.com/steff-farley/AFM-image-segmentation}.
	\end{itemize}

\end{document}